\documentclass[aps,prl,preprint,groupedaddress]{revtex4-2}

\usepackage{amsmath}
\usepackage{amsfonts}
\usepackage{tikz}
\usepackage[shortlabels]{enumitem}
\usepackage{hyperref}

\newtheorem{thm}{Theorem}

\newcommand{\Poi}{\mathrm{Poi}}
\newcommand{\E}{\mathrm{E}}
\newcommand{\critFE}{\rho_c}
\newcommand{\critDD}{\rho_s}

\newcommand{\critthresh}{\rho_t}
\newcommand{\ZZ}{\mathbb{Z}}
\newcommand{\tvnorm}[1]{\lVert #1 \rVert_{TV}}
\renewcommand{\P}{\mathbf{P}}

\begin{document}


\title{Activated random walk exhibits self-organized criticality}


\author{Christopher Hoffman}
 \email{choffman@uw.edu}
 \homepage{https://sites.math.washington.edu//~hoffman/}
 \affiliation{Department of Mathematics, University of Washington}
 \author{Tobias Johnson}
 \email{tobias.johnson@csi.cuny.edu}
 \homepage{https://www.math.csi.cuny.edu/~tobiasljohnson/}
 \affiliation{Department of Mathematics,  CUNY College of Staten Island}
 \author{Matthew Junge}
 \email{matthew.junge@baruch.cuny.edu}
 \homepage{https://www.mathjunge.com/}
 \affiliation{Department of Mathematics, CUNY Baruch College}
 \author{Josh Meisel}
 \affiliation{Department of Mathematics, CUNY Graduate Center}
 \email{jmeisel@gradcenter.cuny.edu}
 \homepage{https://sites.google.com/view/joshmeisel/home}
 
%
 %
 %
%


\date{\today}

\begin{abstract}
 To explain the ubiquity of power laws and fractals in nature,
  Bak, Tang, and Wiesenfeld formulated simple conditions for a system to self-organize into a critical state.
 Dickman, Muñoz, Vespignani, and Zapperi postulated that the self-organized critical state matches
 the critical state in corresponding fixed-energy models undergoing traditional phase transitions. 
   Although the theory has been applied broadly
  over the past five decades, no mathematical model has been proven to exhibit the conjectured behavior.
  Indeed, the originally proposed abelian sandpile model displays nonuniversal behavior
  stemming from its slow mixing. 
  Marking the first result of its kind, we prove that the 1-d activated random walk model mixes quickly into a stationary state 
  with power-law avalanches and limiting critical density that equals the critical value for the fixed-energy version.
\end{abstract}


\maketitle


Bak, Tang, and Wiesenfeld's theory of self-organized criticality (SOC) posited that dynamical systems where energy gradually accumulates and releases in bursts 
can drive themselves without any external tuning to a critical state resembling laboratory phase transitions \cite{BakTangWiesenfeld87}.
They introduced the \emph{abelian sandpile} as a model.
In its \emph{driven-dissipative} form, with particles injected into the interior of the system
and dissipating at the boundary, the model appeared to converge quickly to a state displaying the fingerprints of criticality: self-similarity and power-law observables.
For the abelian sandpile and other related models, 
Dickman, Muñoz, Vespignani, and Zapperi (DMVZ) claimed that this stationary state is identical to the critical state
in \emph{fixed-energy} settings of the models,
which undergo traditional absorbing-state phase transitions \cite{dickman1998self,vespignani2000absorbing}.
From this perspective, self-organized criticality is achieved by a driving mechanism
that pushes the model on its own to a universal critical state.

Fey, Levine, and Wilson presented data refuting this picture for the abelian sandpile in dimension two \cite{fey2010driving,fey2010approach}.
The threshold density for the fixed-energy sandpile and the stationary density for the driven-dissipative
sandpile differ (though only after three decimal places).
This refutation is seen not as a failure of DMVZ's theory, but rather as a
lack of universality for the abelian sandpile model, in that the critical states
of different forms of the model are different. 
As we detail later (and as discussed in \cite{levine2021exact, levine2023universality}),
the primary culprit of this nonuniversality is slow mixing.

The present theoretical foundation of SOC leaves much to be desired. As put by Watkins et al.\ \cite{watkins2016twentyfive}, ``At the theoretical end, none of
even the computer models which are widely accepted as displaying all the hallmarks
of SOC has been solved or even only systematically approximated.
In fact, there is not even a mean field theory that makes any quantitative reference
to SOC taking place in spatially extended systems with some form of boundary at
finite distances.''
In this work, we present rigorous mathematical results that \emph{activated random walk} (ARW), an alternative sandpile
model that evolves stochastically, displays the hallmarks of SOC. Specifically, we prove
that:
\begin{enumerate}[(i)]
\item The critical density for 1-d driven-dissipative ARW exists and equals the critical value for fixed-energy ARW, thus confirming the basic picture laid out by DMVZ.
\item Avalanches in the stationary state are heavy-tailed, a form of criticality also present in the abelian sandpile.
\item Mixing occurs when the driven-dissipative system reaches the critical density, removing a known impediment to universality.
\end{enumerate}
Besides being consistent with Bak, Tang, and Wiesenfeld's original predictions, these results imply
that the system remains in a unique state at the critical density, in contrast to the abelian
sandpile which can be thought of as having multiple critical states. 

Addressing the shortcoming noted by Watkins et al., our findings are the strongest example yet of SOC taking place in a spatially-extended system with boundary at finite distances. We note that unlike the abelian sandpile, ARW is nontrivial in dimension one, with enough complexity for SOC to arise.
After we present our results, we provide the basic ideas behind their proofs. The proofs are carried out in complete detail in the companion papers \cite{hoffman2024density, hoffman2025cutoff}.

\section{Sandpile Models}
The abelian sandpile consists of particles on a given graph, which may contain
sink vertices at which particles accumulate. In the basic dynamics
of the system, a vertex is \emph{unstable} 
if the number of particles present there is greater than or equal to its degree. An unstable
vertex \emph{topples} by sending a particle to each of its neighbors.
A configuration is \emph{stabilizable} if it is possible to make all vertices stable
by successively toppling unstable vertices. For a stabilizable configuration, any order of topplings
resulting in a stable configuration 
will yield the same \emph{odometer} (the function counting how many times each vertex is toppled) and hence the same final configuration.

The \emph{driven-dissipative} version of the model takes place on a box within the $d$-dimensional integer lattice, 
with all vertices outside the box acting as sinks. Starting with some configuration, we add a particle
to a uniformly random vertex in the box and then stabilize the configuration, which
is always possible because of the presence of sinks. Thus we obtain a Markov chain
with a state space of stable configurations on the box, which has a limiting, stationary
distribution. This stationary distribution has an infinite-volume limit as the box size
grows \cite{athreya2004infinite}, and as a consequence there is a limiting stationary density,
known to be exactly $17/8$ for $d=2$ \cite{poghosyan2011return,kenyon2015spanning}.

The \emph{fixed-energy} version of the model takes place on a $d$-dimensional torus
of length~$n$, or on the infinite $d$-dimensional lattice with no sinks.
The \emph{threshold density} for the model on a torus is defined as $\E M / n^d$, where $M$
is the minimum number of particles added (uniformly at random) to the initial configuration
so that it ceases to be stabilizable. On the infinite lattice, the threshold density
is the supremum over $\rho$ such that an i.i.d.-$\Poi(\rho)$ initial configuration
is stabilizable a.s. The data presented by Fey, Levine, and Wilson
show that the limiting threshold density for the fixed-energy abelian sandpile on
2-d tori (and conjecturally, the threshold density on the infinite 2-d lattice)
is approximately $2.125288$, slightly exceeding the stationary density $17/8$.

In activated random walk, particles may be either \emph{active} or \emph{sleeping}.
Active particles move as continuous-time simple random walks. When they are alone
on a vertex, they fall asleep at rate~$\lambda>0$, a parameter of the model.
Sleeping particles do not move, but become active if visited by an active particle.
Just as with the abelian sandpile, driven-dissipative ARW takes place on a box, with sinks along the boundary. By injecting particles uniformly on the box, a Markov chain on configurations of sleeping particles is formed. The fixed-energy version of the model
can again take place on a torus or infinite lattice.
On an infinite lattice, the process \emph{fixates} starting from a given initial configuration
if there is finite activity at each site over all time; otherwise we say that the
process \emph{remains active}. There is a critical value $\critFE$ such that
starting from any initial configuration of active particles
with an ergodic distribution of density $\rho>\critFE$, the system remains active a.s.,
while if $\rho<\critFE$ the system fixates \cite{rolla2012absorbing,rolla2019universality}.

\section{Nonuniversality of the abelian sandpile and its causes}
The observation that the fixed-energy threshold density for the 2-d abelian sandpile
fails to converge to the limiting stationary density $17/8$ prompted an investigation into the cause.
Jo and Jeong attributed the failure to the model's sensitivity to its initial conditions \cite{jo2010comment}
(also studied in \cite{Fey-denBoerRedig05}), which portrays the problem as being of nonuniversality.
Another indication of the model's nonuniversality is its nonspherical limiting
shape when particles are injected at the origin 
\cite{PegdenSmart,levine2016apollonian,levine2017apollonian}.

\begin{figure}\centering
    \begin{tikzpicture}[xscale=3,yscale=2.5]
      \draw[->] (-.03,0)--(4.1,0) node[right] {$\rho$};
      \draw[->] (0,-.06)--(0,2.6) node[above] {$D_\rho$};
      \draw[thick,blue] plot[] file {bracelet.table};
      \foreach \x in {0,.5,...,4}
        \draw (\x,0) -- +(0,-.04) node[below,font=\small] {$\x$};
      \foreach \y in {0,.5,...,2.5}
        \draw (0,\y) -- +(-.03,0) node[left,font=\small] {$\y$};
      \draw (2.475,2.4) rectangle (2.9,2.55);
      \begin{scope}[shift={(2.5,.35)},xscale=4,yscale=80,shift={(-2.475,-2.482)}]
        \draw (2.475,2.482) rectangle (2.9,2.502);
        \draw[thick,blue] plot[] file {braceletzoom.table};      
        \draw[dotted] (2.496608,2.482)--(2.496608,2.502);
        \draw[dotted] (2.48,2.496608)--(2.9,2.496608);
        \draw[dotted] (2.48,2.5)--(2.9,2.5);
        \foreach \x in {2.5, 2.6, 2.7, 2.8,2.9}
          \draw (\x,2.482) -- +(0,-.0005) node[below,font=\small] {$\x$};
        \draw (2.496608,2.502) node[above,font=\small]{$2.4966$};        
        \foreach \y in {2.485, 2.49,2.495,2.5}
          \draw (2.475,\y) -- +(-.01,0) node[left,font=\small] {$\y$};
        \draw (2.9,2.496608) node[right,font=\small] {$2.4966$};
      \end{scope}
      
    \end{tikzpicture}
    \caption{The limiting empirical profile of the density $D_\rho$ of the abelian sandpile
    on the bracelet graph after injection of a density $\rho$ of particles, 
    as rigorously proven by Fey, Levine, and Wilson \cite{fey2010approach}.
    At first glance, it looks like it increases linearly to criticality and then remains there,
    but in fact it grows to a constant approximately equal to $2.4966$
    before slowly increasing toward its asymptotic limit $5/2$.
    The numerical data given by Fey, Levine,
    and Wilson indicate similar behavior for the abelian sandpile on the two-dimensional lattice.
    } \label{fig:bracelet}
\end{figure}
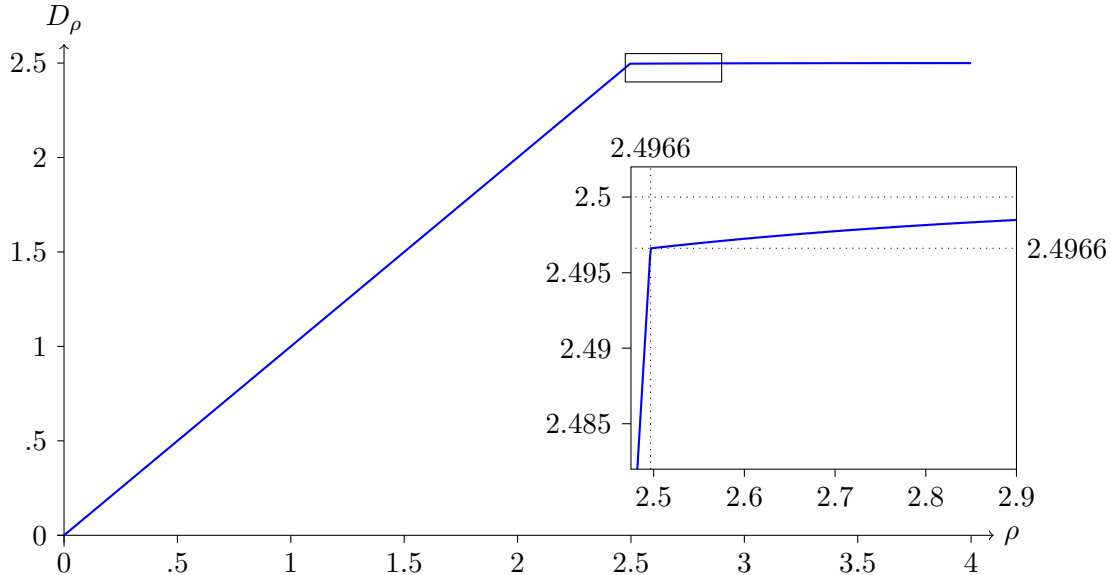

Why is the abelian sandpile nonuniversal?
Bak, Tang, and Wiesenfeld originally suggested that driven-dissipative sandpiles
increase linearly in density until they arrived at criticality, after which
they remain there \cite{BakTangWiesenfeld88}. 
Fey, Levine, and Wilson's work showed this was not so for the abelian
sandpile. On two graphs where the abelian sandpile could be analyzed more easily, the process
increases linearly in density until it reached the fixed-energy threshold, as expected.
But the density continues to evolve after this point, converging slowly to a differing stationary density \cite{fey2010approach}. This behavior points
to slow mixing as the reason the density conjecture fails: when particles are
added to the fixed-energy system, it ceases to be stabilizable before the model has mixed.
An additional finding of Levine's \cite{levine2015threshold}, first conjectured
by Poghosyan, Poghosyan, Priezzhev, and Ruelle \cite{PoghosyanPoghosyanPriezzhevRuelle11}, reinforced this explanation:
For $h<0$, define $\critthresh(h)$ to be the fixed-energy threshold value if one starts
with an initial configuration identically equal to $h$, interpreting negative values as holes
that are filled in by particles. That is, $\critthresh(h)$ is the expected density when the 
configuration first becomes nonstabilizable when adding particles uniformly at random starting from
initial configuration $h$. Then $\critthresh(h)$ converges as $h\to-\infty$ to the stationary
density for the driven-dissipative model. Starting with holes that must be filled in gives the model
more time to mix, allowing the density conjecture to hold again.
The mixing time of the abelian sandpile was later determined very precisely, confirming that
it takes the addition of $\Theta(V\log V)$ particles
to mix on a two-dimensional box of volume $V$
\cite{hough2019sandpiles,hough2021cutoff}. Thus,
the process has not yet mixed in the $\Theta(V)$ steps required to add a critical density of particles.

\section{Prior work on ARW}
The nonuniversality of the abelian sandpile led to interest in sandpile models with stochastic
transitions such as the stochastic sandpile and ARW. Of these two, ARW has been the favored
model since it is somewhat more tractable. Over the past 15 years, mathematicians
have shown the fixed-energy critical density on $\ZZ^d$ is strictly positive
for $d=1$ \cite{rolla2012absorbing} and then for $d\geq 2$ \cite{SidoraviciusTeixeira17,StaufferTaggi18},
and have shown that it is strictly below 1 for $d=1$ \cite{BasuGangulyHoffman18,HoffmanRicheyRolla20},
for $d\geq 3$ \cite{StaufferTaggi18,Taggi19}, and finally for 
$d=2$ \cite{hu2022active,forien2022active,asselah2024critical}.
The importance of fast mixing for universality led Levine and Liang
to investigate the mixing time for ARW \cite{levine2021exact}. They proved that the mixing time for driven-dissipative ARW on a box
of volume $V$ in $\ZZ^d$ is $O(V)$, without the logarithmic term present for the abelian sandpile.
They conjectured that the mixing time is $(\critFE+o(1)) V$, i.e., that ARW has mixed once enough
enough particles have been added to bring it to the fixed-energy critical density.

\section{Results}
We give a rigorous confirmation that ARW in dimension one exhibits universality in various ways
that the abelian sandpile does not. Note that the abelian sandpile has trivial behavior on an interval,
unlike ARW. But even on the  one-dimensional bracelet graph, where the abelian sandpile has nontrivial behavior, the density conjecture fails for it
\cite[Theorem~4]{fey2010approach}.

First, in recent work we have proven that 
the density conjecture holds.
    As before, let $\critFE=\critFE(\lambda)$ be the critical density for fixed-energy ARW on $\ZZ$ with
    sleep rate $\lambda>0$. Let $\critDD^{(n)}=\critDD^{(n)}(\lambda)$ be the expected density under the stationary distribution
  for driven-dissipative ARW on an interval of length~$n$.


  
\begin{thm}
\label{thm:density}
  $\critDD^{(n)} \to \critFE$.
\end{thm}

Next, we confirm Levine and Liang's mixing time conjecture for ARW in dimension one.
To state the result, let $\sigma_0$ denote some initial configuration of particles
on the interval of length~$n$. Let $\sigma_1,\sigma_2,\ldots$ denote successive steps
of driven-dissipative ARW on the interval, i.e., let $\sigma_{i+1}$ be the random
configuration of sleeping particles obtained by adding an active particle to $\sigma_i$ at a uniformly
random location and allowing the resulting configuration to stabilize.
Let $\pi=\pi(n,\lambda)$ denote the stationary distribution of the Markov chain $(\sigma_i)_{i\geq 0}$.
The \emph{total-variation distance} $\tvnorm{\P(\sigma_t\in \cdot)-\pi}$ measures the distance
to stationarity after $t$ steps of the chain; it is the minimal probability
that $\sigma_t$ fails to match $\sigma\sim\pi$ under all couplings of $\sigma_t$ and $\sigma$.
Given $\epsilon \in (0,1)$ and $\lambda >0$, define the \emph{mixing time}
\begin{align*}
  \mathbf{t}_{\mathrm{mix}}^{(n)}=\mathbf{t}_{\mathrm{mix}}^{(n)}(\epsilon,\lambda) := \min\bigl\{t\colon \max_{\sigma_0}\tvnorm{\P(\sigma_t\in\cdot)-\pi} \leq\epsilon\bigr\}.
\end{align*}
Here the maximum is taken over all possible initial configurations $\sigma_0$. The following result holds for all $\epsilon \in (0,1)$ and $\lambda >0$.

\begin{thm}
\label{thm:mixing}
  $\dfrac{\mathbf{t}_{\mathrm{mix}}^{(n)}}{n} \to \critFE$.
\end{thm}

Our result implies that ARW exhibits \emph{cutoff}, i.e., the transition from being far from stationary to being close to stationary occurs over a time window that is negligible compared to the mixing time itself. 

These results portray ARW as very different from the abelian sandpile.
The root cause of nonuniversality for the abelian sandpile, slow mixing, does not
occur for ARW. Indeed, driven-dissipative ARW is mixed as soon as enough particles are added
to bring the density to criticality. Consequently, the critical state is unique, unlike for the abelian sandpile.

\begin{figure}
    \centering
    \begin{tikzpicture}[xscale=9,yscale=4.5]
      \draw[->] (-.03,0)--(1.3,0) node[right] {$\rho$};
      \draw[->] (0,-.06)--(0,1) node[above] {$D_\rho$};
      \foreach \x in {0, .25,.5,.75,.889,1,1.25}
        \draw (\x,0)--(\x,-.06) node[below] {$\x$};
      \foreach \y in {0,.25,.5,.75,.889}
        \draw (0,\y)--(-.03,\y) node[left] {$\y$};
      \draw plot[only marks,mark=*,mark size=.01pt,mark options={blue}] file {hockey2500-2000-.8.table};
    \end{tikzpicture}
    \caption{A sample path of the density $D_\rho$ of driven-dissipative ARW on an interval
    of length~2000 after the addition of a density $\rho$ of particles.
    The critical density $\critFE$ is approximately $.889$.
    The limiting empirical density profile is proven to be $\rho\mapsto\min(\rho,\critFE)$,
    different from Figure~\ref{fig:bracelet}.}
    \label{fig:stick}
\end{figure}
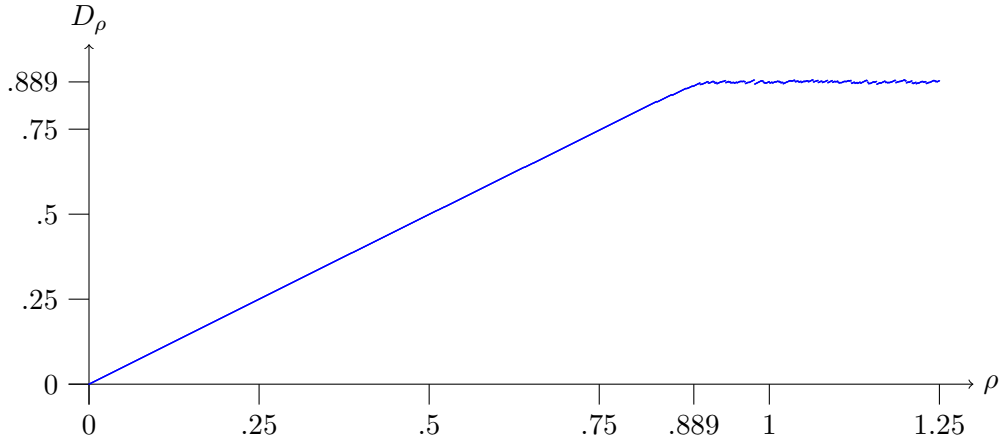

The original behavior predicted by Bak, Tang, and Wiesenfeld was that the sandpile
increases steadily in density until reaching criticality, and then ``[o]nce the critical point
is reached, the system stays there'' \cite{BakTangWiesenfeld88}. As shown
by Fey, Levine, and Wilson (see Figure~\ref{fig:bracelet}), the abelian sandpile
does not follow this description. But ARW adheres to this prediction
(see Figure~\ref{fig:stick}),
as implied by Theorem~\ref{thm:mixing} which states that as soon as enough
particles have been added to bring the ARW system to critical density, it remains there---not
just at the critical density, but in a unique critical state. (The slightly weaker statement
of persisting at the critical density is the \emph{hockey-stick conjecture}
\cite[Conjecture~17]{levine2023universality}. It is an immediate corollary of Theorem~\ref{thm:mixing}
and was also proven in \cite{hoffman2024proof}.)

One might ask what justification there is that the stationary state is indeed critical. 
First, by Theorem~\ref{thm:density} its density matches
the critical density for fixed-energy ARW on the line, at
which a traditional absorbing-state phase transition occurs.
And second, avalanches are heavy-tailed in this stationary state.
To make this precise, let $X=X(n,\lambda)$ denote the \emph{avalanche area} when
adding a particle at the center of the interval at stationarity, i.e., $X$ is the number
of sites that see activity when stabilizing an active particle at the center added
to a configuration of sleeping particles sampled from the stationary distribution $\pi=\pi(n,\lambda)$.

\begin{thm}
$\P(X\geq k/2)\geq k^{-1}$ for $1\leq k\leq n$.
\end{thm}


\section{Proof techniques}

ARW is an abelian model under the sitewise representation, where random jump and sleep instructions
are placed in a stack at each site, and particles move by executing the next instruction of the stack at their current location.
The stabilizing odometer records the number of instructions executed at each site
when stabilizing a configuration. This odometer satisfies mass-balance equations at each site,
and according to the least-action principle is minimal among all odometers satisfying these equations.
Thus, we shift to studying the class of all such odometers. In \cite{hoffman2024density},
we show that this class can be embedded in a stochastic process resembling $(2+1)$-dimensional
directed percolation. Analysis of this process then establishes the density conjecture.

To analyze the mixing time of driven-dissipative ARW, we prove (on arbitrary graphs) what amounts
to the statement that the system has mixed once enough particles have been added that activity is guaranteed at all sites.
The idea is that by a ``preemptive'' abelian property \cite[Lemma~3.1]{hoffman2025cutoff},
sleep instructions can be ignored at a site when future activity is known to occur there.
Thus, if activity is guaranteed everywhere, we may add extra particles to every site without
changing the law of the final configuration, since the added particles may be preemptively toppled to the sink. But the final configuration
after starting with at least one active particle at every site is already known to be exactly 
stationary \cite[Theorem~2.1]{levine2021exact}.
Finally, in dimension one, the machinery of \cite{hoffman2024density} can be used to prove
that any supercritical density is likely to produce activity everywhere.

Our proof that avalanches are heavy-tailed is quite short (and the same argument works
for the abelian sandpile). Starting at stationarity, we produce $k$ sequential avalanches, adding particles to the origin one at a time. Each avalanche results in a stationary configuration, and so the avalanche areas are identically distributed. After all $k$ avalanches are finished, at least one of the added particles must end up at a distance of at least $k/2$ from the origin, and so at least one of the avalanche areas must have been this large. Thus if $P(X\geq k/2)$ were less than $1/k$, by a union bound there would be positive probability that none of the $k$ avalanches have area $k/2$ or larger, yielding a contradiction.

\section{Conclusions and further questions}
Our work demonstrates that 1-d ARW serves as a model of self-organized criticality
with at least some degree of universality, and it supports the DMVZ view
of self-organization as a consequence of a system's natural driving mechanism
toward a critical state. A natural future direction would be to further confirm the DMVZ perspective
by establishing the equivalence of the driven-dissipative and fixed-energy critical states
in more depth. One aspect of this is to
prove local convergence of the driven-dissipative stationary distribution
to a limit that can also be obtained from the fixed-energy or other variants of the model, e.g., as
a limit of final configuration in the point-source model. There are many beliefs
about the structure of this conjectural limit (see \cite[Section~1.3]{rolla2020activated}
and \cite{levine2023universality}) that remain unconfirmed, and there is much else
to determine about the self-organized critical state.
A recent preprint \cite{hoffman2026local} is a first step toward this goal.
Of greater physical importance is extending our work to $d\geq 2$.
Our work is not inherently limited to dimension one, and despite the added difficulty in higher
dimensions, we believe that our odometer-based techniques provide a path toward establishing
a universal model of self-organized criticality in all dimensions.

\bibliography{main}

\end{document}